# Interaction of the ultra-short Bessel beam with transparent dielectrics: Evidence of high-energy concentration and multi-TPa pressure

**Eugene G. Gamaly, Andrei V. Rode**
*Laser Physics Centre, Research School of Physics and Engineering,*
*Australian National University, Canberra ACT 2601, Australia*

**Saulius Juodkazis**
*Centre for Micro-Photonics, Swinburne University of Technology, Hawthorn, Vic 3122, Australia*

**Ludovic Rapp, Remo Giust, Luca Furfaro, Pierre Ambroise Lacourt, John M. Dudley, Francois Courvoisier**
*Institut FEMTO-ST, UMR 6174 CNRS Université de Bourgogne Franche-Comté,*
*25030 Besançon Cedex, France*

E-mail: *eugene.gamaly@anu.edu.au*

## Abstract

It has been proven that the intense tightly focused Gauss beam (GB) generates pressures in excess of a few TPa creating the novel super-dense phases of Aluminium and silicon [1-5]. Recently it was demonstrated that the Bessel beam (BB) focused inside sapphire produced the cylindrical void being two orders of magnitude larger than that generated by the GB [6-8].

Analysis of the experimental data presented below allows making the remarkable conclusions based solely on the void size measurements without any *ad hoc* assumptions about the interaction process. First, the *void size is direct evidence of strong (>40%) absorption* of the pulse energy. *Second, it is a* direct experimental evidence of the high-energy concentration in the central spike of the focus. The unique features of the intense Bessel beam interaction then allow understanding the experimental observation. This interaction generates early in the pulse time the spatial distribution of excited permittivity changing from positive to negative values. Then the light interacts with zero-real-permittivity surface, separating plasma and dielectric areas, which leads to high energy concentration near the axis of cylindrical focus up to several MJ/cm$^3$ (pressure range of 4-8 TPa). The effect depends on the angle between the permittivity gradient and the field polarisation. High pressure generates intense cylindrical shock/ rarefaction waves, which led to formation of void and compressed shell.

We demonstrate that the Bessel beam proves to be an effective tool for producing extreme pressure/temperature conditions on the laboratory tabletop. It appears that adjusting polarisation and permittivity gradient might be a novel way for increasing the maximum pressure. This tool allows for search of novel high-pressure material phases, for the 3D laser machining and for creating Warm Dense Matter as those in stars' cores.

## I. Introduction

It is well known that creation of TPa pressures in the laboratory is a formidable experimental task [9,10]. In stationary conditions the pressures of a few tenth of TPa have been obtained using a diamond anvil cell [10]. The transient pressures up to 50 TPa behind the shock front were generated either by nuclear explosions or with the use of powerful laser [11,12]. These experiments are complicated, cumbersome and very expensive.

Recently TPa pressures, that is larger the strength of any material, and temperatures up to 50 eV were produced using low energy pulses from a conventional tabletop laser having the Gauss spatial distribution of intensity across the beam [2,3]. Moreover the formation of new high-pressure phases of Aluminium and Silicon in these conditions was demonstrated [4,5].

Here we show that it is possible to create extreme conditions in much larger volume using intense non-diffracting Bessel beam. Major difference between the Gauss and BB beam interaction with matter resides in a huge difference in the length of the focal volume: in GB it compares with laser wavelength while the length of the BB focal area is $Z_{max} >> \lambda$ .

It appears that the processes of the formation of intensity distribution in the focal area of the intense BB are strongly correlated with the medium modification by the beam. Early in the interaction of short intense BB with transparent dielectric the generation of conductivity electrons makes the transient permittivity variable in space while the mass density remains intact. The real part of the permittivity changes in space from the positive values in unperturbed dielectric to negative values in dielectric converted to plasma. Plasma appears to be surrounded by the complicated 3D surface where the real part of the transient permittivity is zero.

The electrical inhomogeneity of the laser-affected dielectric and existence of this surface drastically changes propagation/interaction and intensity distribution formation processes. It is possible to show how inhomogeneity may affect the energy concentration during the BB focusing. In what following we present the experimental results on void formation by the BB, and analyse the results. We present the physical picture of the BB interaction with transparent dielectrics accompanied by the calculations and comparing theory with the experimental data. The structure of the paper is as the following. In Section II we present briefly the experimental set up, formulation of the experiments, diagnostic tools and the data obtained, all from the Ref.[8]. In Section III we present the discussion of the experimental data accompanied by the succession of the interaction, estimate absorption, plasma formation, time/space distributions of the laser-modified permittivity of the sapphire converted to plasma. We pay special attention to description of the energy concentration during the BB focusing. Then we consider energy transfer from the heated electrons to ions, building up the pressure, generation, propagation and decay of the shock and rarefaction waves leading to formation of the void. Then we compare the theoretical and experimental results and identify the possible novel features of the BB interaction with the initially transparent dielectrics indicating the differences with the Gauss beam/sapphire interaction. In Section IV we make the conclusions and discuss the future directions.

## II. Experimental

*Laser parameters:* A short and intense Bessel beam (BB) was focused inside sapphire crystal at the depth of a few microns with high NA lens [8]. The laser beams with duration of 140 fs and 3 ps at the energy per pulse up to 2 μJ were created by the CPA technique at the central wavelength 800 nm with 5 kHz repetition rate.

*BB formation*: Pulse duration was measured at sample site. The spatial light modulator (SLM) was used for producing the Bessel beam spatial shaping of intensity within the focal region. SLM was realized along with the telescope arrangement allowing both Fourier filtering and precisely controlling the position of the Bessel beam inside the sapphire samples. The polarization orientation was controlled after Fourier filtering by a half-wave plate. The polarization of the central lobe of the Bessel beam is linear.

*Focussing:* The beam was focused through an Olympus MPLANFLN microscope objective of 50x with 0.8 numerical aperture inside sapphire sample. The position of the sample was controlled by X-Y-Z motorized translation stages and a φ-ϑ motorized translation stage for its flatness (to maintain the sample perpendicular to the laser beam). An independent pulse picker inserted after the laser source controlled single shot operation.

*The sapphire samples* are mono-crystals, C-cut, double-side polished with a thickness of 100 µm. The material and optical properties of sapphire are collected in Appendix 1.

The experimental setup is described in details in [6-8].

*Intensity distribution inside a focus:* It was demonstrated experimentally [6-8] that such a pulse tightly focussed inside sapphire produces long nearly cylindrical void. In order to understand the processes inside a focus of intense laser beam one should have information about intensity distribution inside a focal volume similar to that as it was done with the Gauss shape of the beam [2,3].

The continuous wave (*cw*) low intensity BB was widely studied theoretically and experimentally [6-8,13]. *CW* low intensity BB maintains the spatial distribution of intensity across and along the beam similar to that in the perfect non-diffractive beam. The comparative experiments for propagation of low energy (< µJ) BB in air and sapphire were performed in order to observe the effects of short pulse duration on the spatial structure of intensity in different media.

The Bessel beam was produced with a cone angle of 26 degrees. The focus is ~ 32 µm long in air, the diameter of the central lobe equals to ~ 0.7µm. [8]. The spatial distribution of intensity across the Bessel beam has a structure similar to that of the squared zero-order Bessel function of the ideal non-diffractive beam. The transverse beam profile of a Bessel beam comprises a narrow intensity spike on the beam axis, surrounded by the concentric rings. The relative intensities normalized to that of the central core are the following: 1; 0.21; 0.105; 0.073; 0.042. Note, that the peak intensity of the central part is maintained over approximately 15 µm of propagation distance. In sapphire the general features are similar to those in air.

It is obvious that the fluence (absorbed energy per unit area) exceeds the ionization threshold first in the central spot on the axis. When plasma is formed there the group velocity of the running wave goes to zero and running wave converts into an evanescent wave at the distance of the skin depth, which is usually in the order of several tens of nanometers.

The flow of photons directed to the axis experiences strong absorption and specular reflection from plasma. The process of the diffraction-free Bessel beam formation changes dramatically that we discuss in the next Section.

## A. Experiments at the high energy pulse (~ 2μJ)

Ultrashort and intense single laser pulses shaped to produce a Bessel beam and tightly focused inside the sapphire produce narrow and elongated voids in the bulk of the crystal [6-8]. The intensity of the Bessel beam has maximum at the axis. Therefore, the fluence (absorbed energy per unit area) exceeds the ionization threshold first at the central core close to the axis. At the surface where the ionization threshold is achieved the real part of the modified permittivity goes to zero and the running electromagnetic wave converts into the evanescent wave. The flow of photons at this surface experiences strong absorption and specular reflection from the plasma formed. The process of the diffraction-free Bessel beam interaction with matter changes dramatically from this moment, modifying the spatial distribution of the intensity. Unfortunately, there is no time/space resolved measurements during the pulse and the void formation.

The energy per pulse was 2 μJ at 140 fs and 800 nm [8]. The consequent laser pulses were spatially separated by 20 μm. The focal depth inside the sapphire crystal was changing along definite direction. The threshold for single shot material modification in sapphire was determined by the optical microscopy (NA 0.8, x50 microscope objective) at the laser energy of 1.2 μJ. Appearance of cracks was determined above the 2.4 μJ level and could extend over several micrometers in length. The laser energy was set at 2 μJ per pulse just below the threshold for appearance of cracks for femtosecond pulse duration.

The structure generated by the Bessel beam inside sapphire was studied only by post mortem methods. The samples have been examined using an optical microscope and a scanning electron microscope (FEI Helios 600i). The structures inside the bulk were assessed using a focused ion beam milling. Special care was taken to avoid any possible damage of the laser produced channels by FIB milling [8].

*FIB milling procedure.* The sapphire samples were opened in order to obtain a view of the longitudinal cross-section of the nanochannels. This was performed in three steps. First, FIB milling was performed at high current of 9.3 nA up to a distance of 10 microns from the channels location. Then, the current was decreased from 2.5 nA to 80 pA until a distance of less of a micron from the channel. The final approach was made at the low current of 40 pA to obtain a clean and precise section and stops when the whole nanochannel was clearly seen.

We opened up approximately half way the channel to obtain a full transverse view of the whole length of the channel. The very elongated void obtained at 140 fs, length measured at ~30 μm, with a very narrow and constant width, measured at 300 nm at maximum see Fig.1.

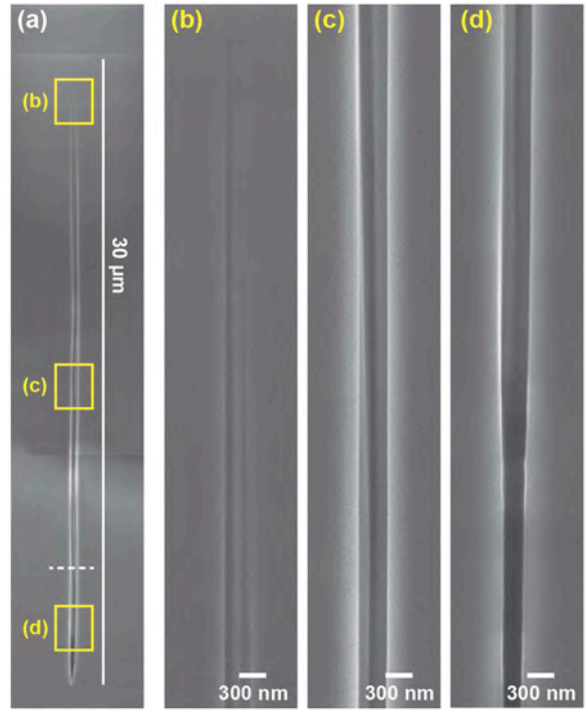

Figure 1. SEM images of a nano-channel formed in sapphire by a single laser pulse shaped as a Bessel beam with the energy of 2 µJ and opened up by focused ion beam milling. (a) – 30-µm long channel; (b, c, d) – are the magnified images at the corresponding sections for demonstration of the width of the channel of ~300 nm. (*Adopted from Ref.[8]*)

## B. Evidence of the high-energy concentration: Analysis of the experimental observations on the basis of laws of mass and energy conservation

Measured void size leads us to the conclusions, which are based solely on the above measurements without any *ad hoc* assumptions about the interaction process. First, the absence of the cracks is the evidence that the material removed from the void volume occurred only due to the cylindrical explosion followed by the generation of strong shock and rarefaction waves similar to the spherical explosion induced by the tightly focused Gaussian beam [2,3]. It means that one can analyse these measurements on the basis of the mass and energy conservation laws. The mass conservation suggests that all the material removed from the void volume remains in the form of compressed cylindrical shell surrounding the void. The minimum work necessary for the removal sapphire from the void of volume $V_{void}$ is $P \cdot V_{void} \geq Y_{sap} \cdot V_{void}$; here $Y_{sap} = 0.4$ TPa = 0.4 MJ/cm$^3$, is the Young modulus of sapphire. Taking the measured average volume of the void, $V_{void} = \pi \times (1.5 \times 10^{-5}\text{cm})^2 \times 3 \cdot 10^{-3}$ cm $= 2.12 \times 10^{-12}$ cm$^3$, one obtains the energy equal to 0.848 µJ necessary to produce the void. Note, that this is a conservative (minimum) estimate not accounting for the thermal energy of the plasma formed, the ionisation losses and the losses of the shock wave energy propagating and compressing the material. Thus, the *void size is the direct evidence of absorption* at least 42.5% of the 2µJ of the pulse energy. Another remarkable result of these measurements is the diameter of the void of 300 nm*, which is 2.5 times less than the diameter of the central spike of the Bessel beam in air*, and close to the diameter of spike in sapphire at low energy (~ 350nm). This is a direct experimental evidence of the increase of the energy concentration in the central spike during the interaction of intense ionizing Bessel beam with the transparent dielectric. It follows from the plasma hydrodynamics that in order to generate a void of volume, $V_{void}$, the volume of the high energy density concentration $V_{pressure}$, where the pressure driving the shock wave is generated, should be much less than that of void, $V_{void} >> V_{pressure}$. In other words, the work for removing material from the void is much higher than that left in $V_{pressure}$ and then consumed by the rarefaction wave [14]. Thus, the measurement of the void size gives immediately the lowest limit of the absorbed laser energy. It is also the evidence of the much higher energy concentration during the interaction of the intense Bessel beam with sapphire than in the interaction of a low intensity beam.

### III. Discussion

#### A. Spatial dependence of the femtosecond BB intensity in the focal region before plasma formation

In what follows we consider the physical reasons for the high energy concentration in the intense Bessel beam interactions. Experiments with the low energy ultra-short Bessel beam presented before show close resemblance to the radial intensity distribution of the ideal *cw* non-diffractive beam. Thus it is reasonable to take the spatial distribution of the electric field (intensity) similar to that in the ideal Bessel beam accounting for the time-dependence of the pulse. Correspondingly, the intensity is taken in the form [13]:

$$I\left(r,z,t\right)=\frac{cE_0^2\left(z\right)}{8\pi}\cdot J_0^2\left(k_r r\right)\cdot\varphi\left(t/t_p\right)=I\left(r,z\right)\cdot\varphi\left(t/t_p\right) \tag{1}$$

Here $J_0$ is the Bessel function of the first kind zero order, $k_z^2+k_r^2=k^2;\ k>k_r>0$, $k$ is the wave number for the field inside a medium. Let's define a width of the central spike as a distance between the Bessel beam zeros on the intensity profile.

The first zero locates at $k_r r_1=2.402$. Taking $k_r\approx k=2\pi n/\lambda$ one gets for the radius of the spike $r_1=2.402\lambda/2\pi\cdot n$ ($n,\ \lambda$ are correspondingly refractive index of a medium and wavelength). For 800 nm in air one gets $r_{1,a}=305$ nm (measured 350 nm), for sapphire ($n=1.75$) $r_{1,s}=175$ nm. One can see a qualitative agreement of these estimates with the experimental figures. Time dependence is taken in the Gauss form:

$$\varphi\left(t/t_p\right)=\exp\left\{-2\left(1-2t/t_p\right)^2\right\} \tag{2}$$

The time dependence of fluence expresses analytically (see Appendix 3):

$$F\left(r,z,t/t_p\right)=\int_0^{t/t_p}I\left(r,z\right)\varphi\left(\tau\right)d\tau=\frac{F\left(t_p\right)}{2}\left\{1+\frac{erf\left[\sqrt{2}\left(2t/t_p-1\right)\right]}{erf\sqrt{2}}\right\} \tag{3}$$

Let's estimate the maximum fluence in the experimental conditions as the following. First, the minimum value one obtains suggesting that the pulse energy arrives at the surface of zero intensity at $r_{1,s}=175$ nm, $L=32$ μm, giving $F=5.68$ J/cm$^2$. As one can see later the ionization and plasma formation thresholds are correspondingly 0.13 J/cm$^2$ and 1.9 J/cm$^2$. Intensity across the central core increases approaching to the axis. It is reasonable to take as an average radius of the surface, where the majority of the incident energy absorbs, the radius where the spatial derivative of intensity is a maximum, which defines by condition, $\dfrac{d^2 J_0^2(k_r r)}{dr^2}=0$. One can find this value from the plot of $J_0^2(k_r r)$ being approximately at $k_r r \approx 1.4$, giving the radius of 102 nm, and the maximum fluence, $F_{max}=9.745$ J/cm$^2$.

### B. Interaction of the Bessel beam at the fluence exceeding the ionization threshold

The interaction of the Bessel beam at the high intensity approaching and overcoming the ionization threshold drastically changes the interference pattern (and intensity distribution) from that at the low intensity. The intensity in the ideal non-diffractive beam at the central spike is five times higher that in the next ring. Therefore the ionization threshold is first achieved near the axis. At this threshold the real part of the dielectric permittivity turns zero, electromagnetic wave is partly reflected from this surface and becomes evanescent below the surface (close to the axis). The cylindrical plasma region shrinks while the intensity of the pulse grows up. The spatial distribution of the absorbed energy density, number density of conduction electrons and the transient permittivity follows the intensity (fluence) spatial distribution making laser and material parameters strongly inter-related. The scheme of the interaction of the Bessel beam with transparent crystal presented at Fig. 2.

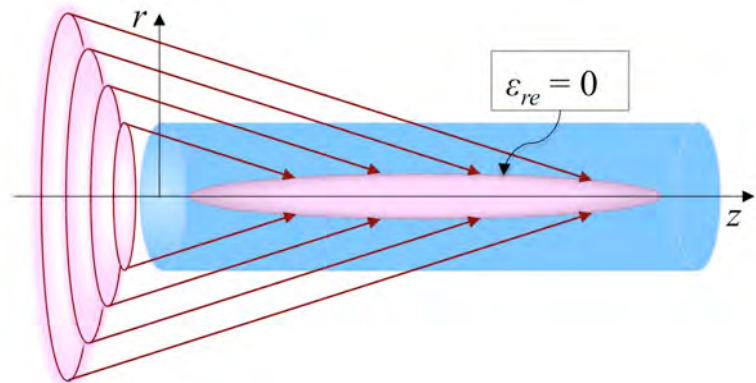

Figure 2. Intense Bessel beam interaction with a transparent dielectric leading to a radially variable permittivity of the media. The central elongated volume indicates the laser induced plasma with the ionisation threshold at the plasma surface where the real part of permittivity $\varepsilon_{re}=0$.

### 1. Transient permittivity in laser-affected transparent dielectric

The approximate expression for the permittivity of the laser-excited dielectric [15] proved to be an adequate approximation for describing space/time dependent optical properties of swiftly ionized transparent solid. Note that the mass (atomic) density remains intact during the time necessary for the energy transfer from hot electrons to the lattice:

$$\varepsilon \approx \varepsilon_0 - \left(\varepsilon_0 - 1\right)\frac{n_e}{n_a} - \frac{n_e}{n_{cr}\left(1 + v^2/\omega^2\right)} + i\frac{n_e}{n_{cr}\left(1 + v^2/\omega^2\right)}\frac{v}{\omega} \qquad (4)$$

Essential feature of this approach is that the dramatic change in the electrons' collision rate during the interaction process is taken into account explicitly. First, electron-phonon scattering is a dominant process. However, in a transparent solid absorption is close to zero due to practical absence of the conductivity electrons. Absorption increases in proportion to the number density of electrons and collision rate. Then the Coulomb collisions rate grows up dramatically approaching the full first ionization state. It depends on the conduction electron number density as it does after the full transformation to the solid density plasma. Thus the permittivity (4) strongly depends on the instantaneous value of the electron number density at any space point. The electrons' number density generated by the multi-photon ionization and by the electrons impact process depends on the space/time dependent absorbed fluence. Spatial pattern of absorbed energy density allows obtaining spatial distribution of excited permittivity.

### 2. Ionization threshold and threshold for full transformation into plasma

The ionization threshold is reached at the moment when the real part of the time-dependent permittivity turns to zero. This condition allows calculating the number density of conductivity electrons using (4) as, $n_{e,ion} \approx \varepsilon_0 n_{cr}$. At this moment the running electromagnetic wave is converted into the evanescent wave. The electric field intensity is exponentially decaying below the zero-surface in direction to the axis of the cylinder.

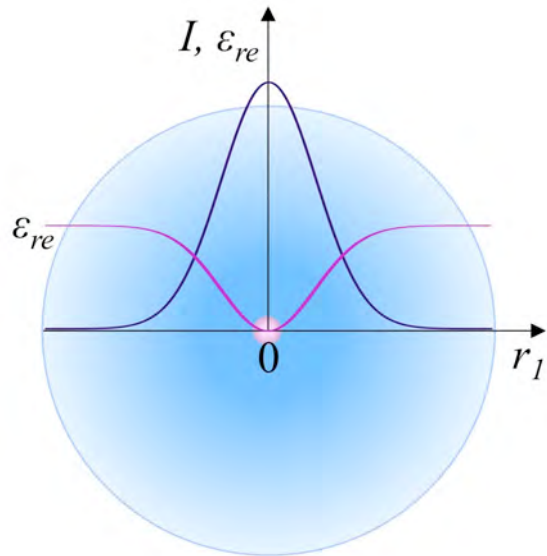

Figure 3. Spatial distribution of laser intensity across the central peak of the Bessel beam and the laser-affected real part of permittivity at the level of intensity when the ionization threshold, $\varepsilon_{re} = 0$, is reached at the maximum of the Gaussian pulse on the bean axis.

The scheme of spatial distribution of the excited permittivity and intensity presented at the Fig. 3. Permittivity is zero not in a point but rather in an area with the scale of electron's mean free path that is of the order of couple of nanometers. The knowledge of the electron number density and assumption that the electron's energy is at the low edge of conduction band, $k_b T_e \approx \Delta_g$ allows to obtain the absorption coefficient and absorption length (distance where the evanescent field decays e-fold).

The ionization threshold is first reached close to the axis because the intensity is a maximum at the cylindrical axis. The plasma area is spreading outside the axis with further increase of intensity forming quasi-cylindrical surface (see Fig. 3).

With the further increase in the absorbed fluence a solid is completely converted to plasma (atomic bonds are broken, each atom is stripped of one electron). The number density of conductivity electrons equals to atomic number density, $n_e = n_a$. At the threshold for complete transformation to plasma the conduction electron energy is a sum of the band gap and the binding energy (energy of cohesion), $k_b T_e \geq \varepsilon_{bind} + \Delta_g$. The fluence for reaching the ionization and plasma threshold [15,16] is obtained from the condition that the absorbed energy is confined in the thermal energy of the electrons with known for both cases electron number density and electrons energy (see Appendix 4 for details). For sapphire the ionization threshold fluence equals to 0.13 J/cm$^2$, absorption length is 165 nm; the plasma threshold fluence is ~ 1.9 J/cm$^2$, and absorption length is 29 nm. Thus the incident energy absorbs in a cylindrical ring and electric field transmitted through this ring goes in direction of the cylinder axis passing the distance much less that the field wavelength. However, absorption and reflection now occurs in the vicinity of zero-permittivity point in electrically inhomogeneous medium that significantly influences the interaction process.

### 3. Interaction of the incident BB in vicinity the zero-permittivity-surface ($\varepsilon_{re} = 0$)

Forsterling found [17,19] that $\varepsilon_{re} = 0$ might be a singular point for the Maxwell equations in the non-absorbing medium. Indeed, the Maxwell equations converted to the single equation for the magnetic field read:

$$\Delta H - \frac{\varepsilon}{c^2}\frac{\partial^2 H}{\partial t^2} + \frac{1}{\varepsilon}\nabla\varepsilon\times\left(\nabla\times H\right) = 0 \qquad (5)$$

Thus, the last term in equation (5) might be infinite at zero permittivity if the numerator in this term is non-zero that depends on the filed polarisation. Forsterling found the solution for the Maxwell equations where the fields are logarithmically diverging close to the zero permittivity point. Gildenburg indicated later on [18,19] that even infinitesimally small absorption kills the singularity leaving the sharp increase in the amplitude of the fields near the point where the real part of permittivity is zero. Moreover, there is the energy flux from dielectric to the plasma-like medium as the real part of permittivity of plasma is negative [18,19]. For s-polarisation case (the field is perpendicular to the plane of incidence) the increase in the amplitude is small in accord to the solution of the Airy equation [19].

The significant increase in the field amplitude near zero permittivity occurs when there is the electric field component along the permittivity gradient (oblique incidence). Qualitatively such increase near the zero-real-part-permittivity (ZPP) is seen from the Maxwell equations [19]. Indeed, keeping only the fast component of the field one gets that $E \propto \varepsilon^{-1}$. Thus the field amplitude in the vicinity of $\varepsilon_{re} = 0$ increases as $E_{max}/E_0 = \left(\dfrac{\varepsilon_0}{\varepsilon_{im}}\right)$. Correspondingly, the intensity increases as $I_{max}/I_0 = \left(\dfrac{\varepsilon_0}{\varepsilon_{im}}\right)^2$. Note, that absorption strongly depends on the effective collision rate, $\varepsilon_{im} \propto n_e \cdot \nu_{eff}$. Conservative (minimising) estimate for sapphire gives about 10 times intensity increase when compared with that in the incident wave.

### 4. Spatial profiles of the intensity and permittivity

The qualitative picture for the spatial distribution of the intensity (fluence) and permittivity across the beam axis after the ionization threshold arises from the above considerations are shown in Fig. 4.

The spatial derivatives of permittivity, number density of the conductivity electrons and intensity are interconnected, as it can be seen from (4). The gradient of the real part of the permittivity is large and negative in accord with the positive gradient of the electrons number density.

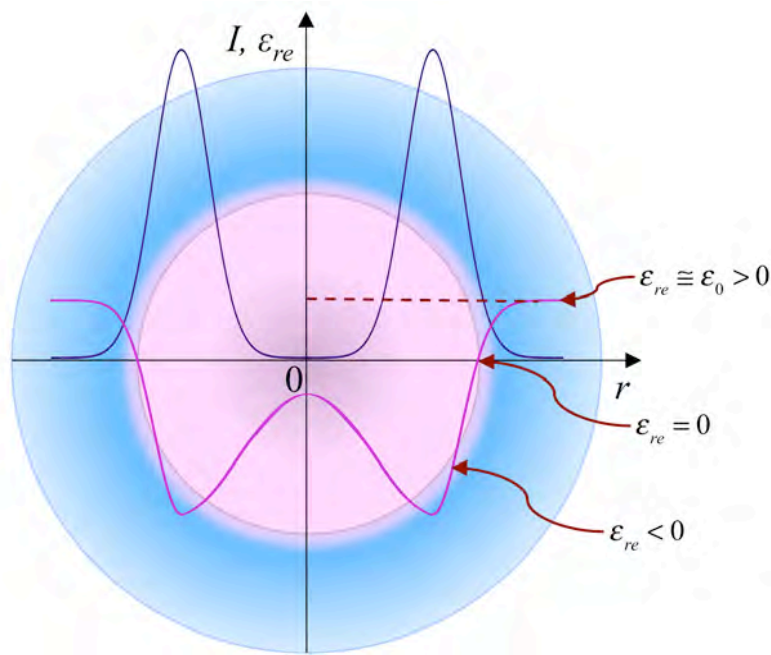

Figure 4. Profiles of the absorbed laser fluence (black line) and of the real part of permittivity (purple line) across the Bessel beam induced plasma. The central area (in purple) shows the region where the electron density is above the ionisation threshold and the real part of permittivity is negative $\varepsilon_{re} < 0$.

### 5. Stability of the Bessel Beam focussing

The negative real part of the permittivity reaches minimum at the maximum of intensity near the $\varepsilon_{re} = 0$ point (see Fig.4). In the evanescent wave deeper into the plasma the laser intensity decreases, the generation of conductivity electrons goes down diminishing the absolute value of the negative $\varepsilon_{re}$. This converts into a positive spatial gradient of $\varepsilon_{re}$ inside the dense plasma as can be seen in Fig. 4

Thus, $\nabla \varepsilon_{re}(r) < 0$ when $\nabla n_e(r) > 0$, while $\nabla n_e(r) \propto \nabla I(r)$. One can see that when the laser intensity goes up, the number density of electrons increases, along with the negative part of the permittivity. Further down, the gradient of $\varepsilon_{re}$ becomes positive when the intensity decreases along with electron number density in direction toward the beam axis. Note that in the conventional low-intensity optics the positive gradient of $\varepsilon_{re}$ means that the electro-magnetic wave propagates from the lesser optically dense media to the higher density medium. In self-focusing situation the increasing of the beam intensity to the centre of the beam leads to the increase in the permeability and decrease in the phase velocity, finally resulting in the collapse of the wave (self-focusing). On the other hand, the self-focusing effect is the consequence of the instability of the wave in the conditions when permittivity grows up along with increasing intensity [19]. Note also that such instability develops when the energy flux through the beam cross section exceeds some threshold. Due to the intensity increase near the $\varepsilon_{re} = 0$ point and correspondingly growth of absorption, the energy flux continuously increases, and that may results in the instability of the wave propagation. However, one should bear in mind that in the case of focussing of the intense Bessel beam all these phenomena occur with the evanescent waves. One may expect a qualitatively similar behaviour of the evanescent wave as it happened with the running wave. However, quantitative conclusions on the stability and the collapse of the evanescent wave can only be made after the studies of the stability problem.

### 6. Estimation for the diameter of the energy deposition region and the absorbed energy density

Analysis above shows that ionization threshold occurs at the fluence of 0.13 J/cm$^2$ at the radius of the central core of 100 nm. At the point of $\varepsilon_{re} = 0$ a significant, up to 10 times, increase in the intensity (fluence) occurs creating a significant increase in the energy flow in the $z$-direction of the cylindrical axis of the beam and thus acting as a focusing lens. The focusing of the evanescent waves to the axis rises the energy density in the region. One might suggest conservatively that the minimum radius of the central region is of the order of magnitude of the absorption length in the solid density plasma that is about 30 nm. Then, taking the minimum absorption coefficient deduced from experiments $\geq 0.4$, one gets the energy density (pressure) in the cylindrical volume with the radius 30 nm and of the length of 32 μm, which is approximately ~9 MJ/cm$^3$ = 9 TPa.

### C. Electronic heat conduction, energy transfer from electrons to ions, shock/rarefaction wave generation and propagation

The non-linear electronic heat conduction starts after the generation of sufficient amount of electrons gaining energy by absorption. Electron diffusion becomes noticeable around the ionization threshold. The electrons' diffusion coefficient $D = v_e^2 / 3\nu$ at the maximum collision rate reads $D = 2.67$ cm$^2$/s. Thus, the electron heat diffusion length at the end of the laser pulse is $l_{heat} = (D \cdot t_p)^{1/2} = 6$ nm, indicating that the heat conduction during the pulse time is still insignificant. The energy transfer rate from electrons to ions expresses through the momentum transfer rate as the following: $\nu_{en} = m_e \nu_{mom} / M$ [20]. Taking the average ion mass for sapphire $M_{av} = 20$ a.u. one obtains the energy transfer rate of $\nu_{en} = 1.37 \times 10^{11}$ s$^{-1}$, and the energy transfer time of 7.3 ps. The heat conduction during the electron-to-ion transfer time becomes significant. Indeed, the heat conduction length at the end of the energy transfer constitutes 44 nm, giving the space scale for the region where the pressure driving the shock wave builds up. The average pressure in a cylindrical region (radius $r_{abs}$ = 44 nm, length 30 μm) is around 4 TPa assuming minimum 40% absorption. Thus, the maximum pressure ranges are 4-8 TPa, which is in agreement with the conclusions from the void measurements. The hydrodynamic motion starts when ions gain the energy from the electrons. Thus 7 ps after the end of the laser pulse the cylindrical shock wave is generated at the radius of ~ 40 nm from the axis. Note that the radius of the energy deposition region, ~ 40 nm, is much smaller than the length of cylindrical focus, which is 30 μm. Thus, assumption of the nearly cylindrical explosion holds well until the distances are comparable to the radius at the both ends of the nearly cylindrical focus. This assumption is in a good agreement with the measurements of the void (see Fig.1).

The formation of the void and the shock-affected zone by the Bessel beam can be understood from the laws of mass and energy conservation similarly to the interpretation of nearly spherical micro-explosion produced by the Gauss pulse in sapphire [2,3,14]. We consider cylindrically symmetric motion. The shock wave loses its energy while propagating through a cold material due to the dissipation in the course of the work done against the internal pressure (Young modulus) that resists the material compression. The distance at which the shock front effectively stops defines the shock-affected volume. At the stopping point the shock wave converts into a sound wave, which propagates further into the material without inducing any permanent changes. The distance where the shock wave stops can be estimated from the condition that the internal energy in the volume inside the shock front is comparable to the absorbed energy, $\pi r_{stop}^2 L \cdot Y = \pi r_{abs}^2 L \cdot P_{max}$. In other words, at this position the pressure behind the shock front equals to the internal pressure of the cold material, i.e. the Young modulus [2,3,14,21]. One can see that the stop distance expresses as, $r_{stop} = r_{abs} \left( P_{max} / Y \right)^{1/2}$. Taking $P_{max} \approx 4$ TPa and $Y_{sapphire}$ = 0.4 TPa one gets $r_{stop}$ = 139 nm.

Void formation inside a solid is only possible if the mass initially contained in the volume of the void was pushed out and compressed. Of course, this is only possible if material escape through the cracks is excluded. After the micro-explosion the whole mass initially confined in a void volume pushed out and compressed into the cylindrical layer with thickness of Δr and the length of the cylindrical void L. Now, one can apply the mass conservation law to estimate the density of the compressed material in the cylindrical shell surrounding the void:

$$\pi r_{void}^2 L \cdot \rho_0 = \pi \left[ \left( r_{void} + \Delta r \right)^2 - r_{void}^2 \right] \cdot L \cdot \rho_{comr} \qquad (6)$$

Thus degree of compression of the material is connected to the radius of the void and the thickness of the shell containing material expelled from the void:

$$\frac{\rho_{comr}}{\rho_0} = \frac{r_{void}^2}{\left( 2 r_{void} \Delta r + \Delta r^2 \right)} \qquad (7)$$

It is clear from Eq.(7) that the thickness of the compressed ( $\rho_{comr} > \rho_0$ ) shell is less than half of the void radius when degree of compression is 1.1 -1.15. Thus the measured void radius is a direct indication on the amount of laser-affected material contained in the shell. The total distance where the hydrodynamic motion, which occurs before the transformation to the sound wave, thus equals approximately one and a half of the void radius, 1.5 $r_{void} \approx 225$ nm (taking the measured void radius length of 150 nm). It is also possible to estimate the duration of hydrodynamic stage taking the average speed of the material expansion as $(Y/\rho_0)^{1/2} \sim 10^6$ cm/s, giving the time of expansion of ~20 ps.

The authors of Ref. [22] produced series of open-ended cylindrical holes with the powerful ultrashort zero-order Bessel beams. They weighted the sample before and after the experiments and found that the mass is practically unchanged. The preservation of mass means that in the Bessel beam induced explosion the shock-wave transfers almost all the material inside the void onto the cylindrical walls. This finding is understandable and compatible with the mechanism explained above. The shock propagation time across the 300 nm wide cylinder is much shorter than the time of the possible outflow through the open ends. The cylindrical symmetry of explosion is broken at the cylinder length comparable to the diameter. Thus the mass loss through the ends is of a few percent, as it was measured in [22].

The same authors presented at the conference in China attended by one of the authors of this paper (S.J.) the measurements of the average speed of propagating shock wave being 40 km/s that corresponds to the driving pressure $P \approx \rho_0 v^2 = 6.4$ TPa also compatible with the results presented here.

### D. Laser-affected material return to the ambient conditions

After the shock wave stopping the material is still hot and continue cooling down by the means of conventional heat conduction of sapphire until the room temperature. Taking conventional equilibrium parameters of sapphire (Dulong Petit heat capacity and heat conduction) one obtains that a material cools down to the room temperature at the distances of few microns from the focal area during a period of a few microseconds. Thus one may expect that shock and heat affected sapphire locates at the distances around 2-3 hundred nanometers from the focal/ energy deposition area.

### IV. Conclusion

The presented experimental data and theoretical analysis of the intense Bessel beam (2 μJ, 140 fs) interaction in the bulk of sapphire allow drawing the following conclusions.

First, the BB tightly focused inside sapphire produced the cylindrical void two orders of magnitude larger than that generated by the GB [6-8]. The void was open by the FIB technique and then carefully measured under SEM. These measurements and their analysis allowed to identifying the consecutive stages of the intense BB interaction with a transparent dielectric. We demonstrate the novel features of this interaction in comparison with those of the tightly focused Gauss beam and low intensity BB. First, the focal region of the BB comprises several tens of microns that is almost hundred times longer the sub-micron focal spot of tightly focussed Gauss beam. Therefore, the build up of the whole focal area continues during the pulse. Second, the intensity in the central spike approximately 5 times larger then that in the next ring. Therefore conversion to plasma at the ionization threshold first occurs close to the central axis in the beginning (around 20 fs) of 140 fs pulse. The cylindrical surface of zero-permittivity, separating plasma and dielectric areas, is created. The absorption and reflection of the incident light occurs at this surface changing the spatial pattern of the incident rays interference and therefore the spatial distribution of the field in comparison to that in the low intensity non-diffracting beam. Analysis shows that zero-permittivity point, ZPP, is a specific point on the permittivity gradient where intensity of the incident light increases several times. The energy flow in the direction to the axis grows up thus increasing the absorbed energy density. The permittivity and intensity gradients in the absorption region have relations similar to those leading to the self-focusing that also results in the energy density increase. We estimate the energy deposition region and found that maximum pressure is in a range of 4-8 TPa.

After the energy transfer from electrons to ions in 7 ps time the strong shock wave accompanied by the rarefaction wave is generated finally leading to the formation of void size in agreement with the observations. Analysis similar to that in [2,3,21] shows that shock and heat affected material confined in a shell with the thickness of ~ 70 nm. This material was converted to high temperature (~ $10^6$ K) solid density plasma with extreme heating rate of ~$10^{18}$K/s and with the cooling rate of ~$10^{16}$K/s. One may expect unusual phase transformations in these conditions similar to those found at the heating by the Gauss beam. Note that amount of material subjected to the extreme conditions by the BB is almost two orders of magnitude higher than transformed by the Gauss beam.

The future studies improving our understanding and control over the BB/ transparent dielectric interaction might be pump-probe studies with the sub-picosecond time resolution. Indeed the shock wave propagates during 20 ps after the end of the pulse. Measuring shock front velocity will allow deducing the driving shock pressure complementing the conclusions from the void size measurements. First steps in this direction were made in attempts of time resolved measurements of the BB produced drilling from the surface [22].

Summing up we demonstrated that the Bessel beam proves to be an effective and controllable tool for producing extreme pressure/temperature conditions in transparent materials on the laboratory tabletop. This tool allows for search of novel high-pressure material phases, for the 3D laser machining and for creating Warm Dense Matter as those in stars' cores.

## Acknowledgements

We acknowledge funding from the Australian Government through the Australian Research Council's Discovery Projects funding scheme (DP170100131) as well as the European Research Council (ERC) under Horizon 2020 program (GA N° 682032-PULSAR).

### Appendix 1: Sapphire parameters

- Molecular mass $Al_2O_3$: 102 a.u. (20.4 a.u. per atom in average)
- Density: 3.95 – 4.1 g/cm$^3$
- Atomic number density: $n_a = 1.2 \times 10^{23}$ cm$^{-3}$
- Refractive index at 0.8μm: $n = 1.75$ (epsilon = 3.13)
- Band gap: 7.3 eV
- Binding energy Al-O: 512 kJ/mol or approx. 5.5 eV
- Ionisation potential: $J_{Al} = 6$ eV; $J_O = 13.6$ eV; average $J_{Al-O} = 10$ eV
- Maximum collision rate ~ $5 \times 10^{15}$ s$^{-1}$
- Young modulus 0.4 TPa
- Thermal conductivity 30 Wm$^{-1}$K$^{-1}$; D = 0.06 cm$^2$/s
- Melting point 2,345 K; boiling point 3,250K

### Appendix 2: Time dependence of the fluence for the Gauss pulse

$$F\left(r,z,t/t_p\right) = \int_0^{t/t_p} I\left(r,z\right)\varphi\left(\tau\right)\cdot d\tau = I\left(r,z\right)F\left(t/t_p\right)$$

For the Gauss pulse the fluence has analytical expression:

$$\frac{F\left(t/t_p\right)}{F\left(t_p\right)} = \frac{1}{2}\left\{1 + \frac{erf\left[\sqrt{2}\left(2t/t_p - 1\right)\right]}{erf\sqrt{2}}\right\}$$

$$F\left(0\right) = 0; \; F\left(t_p\right) = t_p\sqrt{\frac{\pi}{2}}\frac{erf\sqrt{2}}{2} = 0.598 t_p$$

Here $erf\left(y\right) = \dfrac{2}{\sqrt{\pi}}\displaystyle\int_0^y e^{-x^2}dx$ is the error function (erf (2) = 1, $erf\sqrt{2} = 0.9545$).

### Appendix 3: Electrons' momentum exchange rate: electron-phonon and Coulomb collisions

Electrons transferred to the conduction band exchange momentum in collisions with phonons when the electron density is low. At some electrons number density the Coulomb collisions with electrons and ions become dominant. Let's consider both processes.

Electron-phonon collision rate at the lattice temperature **$T$** estimates as the following:

$$\nu_{e-ph} \approx C_{ph}\frac{k_B T}{\hbar}$$

Here $C_{ph} > 1$ is the numerical coefficient introduced to match the experimental value when available. At the room temperature $k_B T = 0.025 eV$ one gets $\nu_{e-ph} \approx 3.8 \times 10^{13}$ s$^{-1}$.

The electron-ion collision rate depends on continuously growing the electron number density and electrons' velocity, which might be approximately considered being practically constant during the ionization process and equal to $v_e \approx (2\Delta_g/m_e)^{1/2} = 2\times10^8$ cm/s. The electron-ion collision rate reads [21]:

$$\nu_{e-i} \approx 3\cdot10^{-6} \ln\Lambda \cdot \frac{n_e Z}{\theta_{e,eV}^{3/2}}; \ \Lambda = 9N_D / Z; \ N_D = 1.7\cdot10^9 \left(\frac{\theta_{e,eV}^3}{n_e}\right)^{1/2}$$

We take $Z=1$, $\theta_{e,eV} = 10$ eV; $\varepsilon_0 =3.13$. The collision rate grows up to the natural maximum, which is defined by the ratio of the electrons' velocity to the inter-atomic distance and comprises ~$5\times10^{15}$ s$^{1}$ (or ~$10^{16}$ s$^{-1}$) [21]. At $\varepsilon_{re} = 1$, (at 800 nm $\omega =2.356\times10^{15}$s$^{-1}$) $n_{cr} = m_e c^2\pi / e^2\lambda^2 = 1.17\cdot10^{13} / \lambda^2 = 1.828\times10^{21}$ cm$^{-3}$); $n_e = 3.89\times10^{21}$ cm$^{-3}$; $N_D = 0.86$; $\ln\Lambda = 2.05$; $\nu_{e-i} = 0.75\times10^{15}$ s$^{-1}$.

It is reasonable using for the description of the collision rate near the maximum the following interpolation [21]:

$$\nu_{eff}^{-1} = \nu_{e-ph}^{-1} + \nu_{e-i}^{-1}$$

Thus near epsilon 1 point the electron-phonon collisions dominate.

### Appendix 4 : Ionization threshold and solid plasma formation thresholds

#### A4.1. Ionisation threshold for sapphire [15]

One can find the threshold flunce corresponding to the ionization threshold from condition that all absorbed energy density confined solely in the thermal energy of electrons transferred to the conduction band.

Critical density at 800 nm $(2.356\times10^{15}$s$^{-1})$ $n_{cr} = m_e c^2\pi / e^2\lambda^2 = 1.17\cdot10^{13} / \lambda^2 = 1.828\times10^{21}$ cm$^{-3}$; electron density at the threshold, $n_{\varepsilon_{re}=0} \approx \varepsilon_0 \cdot n_{cr} = 5.598\times10^{21}$ cm$^{-3}$; electrons collision rate, $\nu_{coll} \approx 2\cdot10^{-7}(cm^3 / s)\cdot n_{e,th} = 1.12\times10^{15}$ s$^{-1}$. One immediately calculates the imaginary part of the permittivity, $\varepsilon_{im} = \dfrac{\varepsilon_0 \nu}{\omega\left(1+\nu^2 / \omega^2\right)} \sim 1.19$; refractive index, $n = \kappa = \left(\varepsilon_{im} / 2\right)^{1/2} = 0.77$; and absorption coefficient $A = \dfrac{4\kappa}{\left(\kappa + 1\right)^2 + \kappa^2} = 0.83$

$$F_{i,th}\left(t_{i,th}\right) = \frac{3c \cdot n_{e,th}\Delta_g}{4A\left(n_{e,th}\right)\cdot\omega\cdot\kappa\left(n_{e,th}\right)} = 0.13 \text{ J/cm}^2$$

One can see that low-density plasma at the ionization threshold ($n_e \ll n_a$) is generated first at maximum intensity in spike early in the pulse time. E-fold decrease length of the evanescent wave in sapphire at the ionization threshold:

$$l_{skin} = c / \kappa\omega = 165 nm.$$

#### A4.2 Threshold for formation solid density plasma ($n_e = n_a$) in sapphire (ablation for a free surface [16])

In conditions, $n_e = n_a = 1.21\times10^{23}$ cm$^{-3}$ ; $\nu = \nu_{max} = 5\times10^{15}$ s$^{-1}$, one gets:

$$\varepsilon_{im}^{a} = \frac{n_a}{n_{cr}\left(1+\nu_{max}^2/\omega^2\right)}\frac{\nu_{max}}{\omega} = 25.55; \quad \varepsilon_{re}^{a} \approx 1 - \frac{n_a}{n_{cr}\left(1+\nu_{max}^2/\omega^2\right)} = \text{-}11.06$$

Then refractive indices obtained from:

$$k_a^2 = \frac{-\varepsilon_{re}^{a} \pm \left[\left(\varepsilon_{re}^{a}\right)^2 + \left(\varepsilon_{im}^{a}\right)^2\right]^{1/2}}{2}; \quad 2nk = \varepsilon_{im}^{a}$$

Real $n_a$ =2.9, and imaginary part of refractive index, $k_a = 4.41$; $R = \dfrac{\left(n-1\right)^2 + k^2}{\left(n+1\right)^2 + k^2} = 0.665$; absorption coefficient, $A_a = (1\text{-}R) = 0.335$, and skin depth $l_a = 29$ nm.

We suggest that electron's energy at this threshold is $k_B T_e \approx \Delta_g + \varepsilon_b$. Then the threshold fluence obtained from condition that absorbed energy density is confined in conduction electrons:

$$F_a\left(t_{th}\right) = \frac{3c \cdot n_a\left(\Delta_g + \varepsilon_b\right)}{4 A_a \cdot \omega \cdot \kappa_a} = 1.9 \text{ J/cm}^2 .$$

Thus, the energy density at the plasma formation threshold estimates as $F_a/l_a = 0.65$ MJ/cm$^3$.